\documentclass[nofootinbib,preprintnumbers,twocolumn,amsmath,amssymb,showpacs]{revtex4}

\usepackage{graphicx}
\usepackage{slashed}
\usepackage{bbm}
\usepackage{footmisc}
\usepackage{multirow,hyperref}
\usepackage{breakurl} 
\usepackage{color}

\renewcommand{\thefootnote}{\fnsymbol{footnote}}
\newcommand{\rig}{\rightarrow}

\newcommand{\be}{\begin{eqnarray*}}
\newcommand{\ee}{\end{eqnarray*}}
\newcommand{\gl}[1]{(\ref{#1})}

\newcommand{\bee}{\begin{eqnarray}}
\newcommand{\eee}{\end{eqnarray}}
\newcommand{\beeq}{\begin{equation}}
\newcommand{\eeeq}{\end{equation}}
\newcommand{\gev}{~{\text{GeV}}}

\renewcommand{\vec}{\bf}

\newcommand{\fb}{~{\text{fb}}}
\newcommand{\ifb}{~{\text{fb}}^{-1}}

\begin{document}

\title{Ditau jets in Higgs searches}

\begin{abstract}
  Understanding and identifying ditau jets -- jets consisting of
  pairs of tau particles, can be of crucial importance and may even
  turn out to be a necessity if the Higgs boson decays dominantly to
  new light scalars which, on the other hand, decay to tau pairs.  As
  often seen in various models of BSM such as in the NMSSM, Higgs
  portals etc., the lightness of these new states ensures their large
  transverse momenta and, as a consequence, the collinearity of their
  decay products.  We show that the non-standard signatures of these
  objects, which can easily be missed by standard analysis techniques,
  can be superbly exploited in an analysis based on subjet
  observables.  When combined with additional selection strategies,
  this analysis can even facilitate an early discovery of the Higgs
  boson. To be specific, a light Higgs can be found with $S/\sqrt{B} \gtrsim
  5$ from $\mathcal {L} \simeq 12~\ifb$ of data. We combine all these
  observables into a single discriminating likelihood that can be
  employed toward the construction of a realistic and standalone
  ditau tagger.
\end{abstract}

\author{Christoph Englert} \email{c.englert@thphys.uni-heidelberg.de}
\affiliation{Institute for Theoretical Physics, Heidelberg University,
  69120 Heidelberg, Germany}
\author{Tuhin S. Roy} \email{tuhin@u.washington.edu}
\affiliation{Department of Physics, University of Washington, Seattle,
  WA 98195, USA}
\author{Michael Spannowsky} \email{mspannow@uoregon.edu}
\affiliation{Institute of Theoretical Science, University of Oregon,
  Eugene, OR 97403, USA}

\pacs{12.60.Fr, 13.85.-t, 14.60.Fg}

\maketitle

%%%%%%%%%%%%%%%%%%%%%%%%%%%
%%%%%%%%%%%%%%%%%%%%%%%%%%%
%%%% SECTION
%%%%%%%%%%%%%%%%%%%%%%%%%%%
%%%%%%%%%%%%%%%%%%%%%%%%%%%
\section{Introduction}
\label{sec:intro}

Revealing the mechanism of electroweak symmetry breaking (EWSB) is the
primary goal of the LHC experiment. The Higgs boson, being a relic of
EWSB, provides a phenomenological window through which 
to study the anatomy of the
symmetry breaking sector.  The discovery of the Higgs boson along with
the determination of its interactions through the measurement of its
mass, spin, ${\cal{CP}}$-charge, branching ratios and production rates
(see, for example \hbox{Refs.~\cite{Gao:2010qx,
    Aad:2008zzm,Ball:2007zza})} are, thus, of paramount interest.
Decades of extensive theoretical and experimental research have
ensured that the discovery potentials of the Higgs in the diboson
channels ($h \rig ZZ/W^+ W^-/\gamma \gamma$) are well
understood. Recent advancement in the jet substructure techniques have
resurrected the $h \rig b\bar b$ channel when the Higgs is boosted
\cite{Butterworth:2008iy,Soper:2010xk}. The significance of this
development lies in the fact that the $b \bar b$ mode is the dominant
decay mode for a light SM Higgs and this channel is shown to be more
potent than the conventional channels for extracting the Higgs boson
from new physics event samples~\cite{MSSM, Tprime}.  These are also
nicely supplemented by the searches in the ditau decay channel of the
Higgs.  Since both the CMS and the ATLAS are expected to identify
tau-jets with reasonable tagging efficiency~\cite{exitau}, the $h \rig
\tau^+ \tau^-$ channel is also considered to be a discovery
mode~\cite{Aad:2008zzm,Ball:2007zza,Plehn:1999xi} for the Higgs boson.

The non-standard cases where the Higgs dominantly decays to four or more
particles via intermediate light resonances are, on the other hand,
relatively less well studied.  Examples of such decays are prevalent
in models of theories beyond the SM, where new scalar or vectorial
degrees of freedom often play the role of the intermediate light
particle.  While a vector boson typically couples uniformly to all
generations of SM particles and thus gives rise to seemingly easier
decays such as $h \rig 4 \ell$ ($\ell = e, \mu$) or slightly more
challenging $h \rig 4 j $~\cite{Martin:2011pd}, a scalar usually
couples maximally to the fermions in the third generation and
facilitates $h \rig 4 b$ and $h \rig 4 \tau$ decays.  The four tau
decay modes can be even more crucial when the intermediate scalar is
too light to decay to $b \bar b $.  The most well known candidate for
an intermediate scalar of these properties is the light pseudoscalar
$a$ in the next-to-minimal supersymmetric SM or NMSSM \cite{Dermisek:2005ar} 
(for a review see Ref.~\cite{Maniatis:2009re}).  Other examples include the light
pseudo-Nambu-Goldstone bosons in theories with dynamical EWSB
(cf. Refs.~\cite{Giudice:2007fh,compsearch}) and even hidden sector
scalars in models of Higgs
portals~\cite{Schabinger:2005ei,hiddenpheno}.

The object of this paper is to investigate the discovery potential of
the Higgs in the following decay channel:    
\begin{equation}
  \label{eq:ditaudec}
  h \, \rig  \, \mathcal{A}\mathcal{A} \, \rig \, \left( \tau^{+}\tau^{-} \right)
  \left( \tau^{+}\tau^{-} \right) \,, 
\end{equation}
where $\mathcal{A} $ represents a light scalar that decays only to a
pair of tau particles. Analyzing this channel is, however, nontrivial
with the existing tau-taggers. Being light, these ditau resonances are
almost always boosted and consequently, often result in a single jet
consisting of a pair of tau particles (dubbed a ``ditau jet'') that
fails to be identified as a tau-jet, an isolated lepton, a photon, or
even a ``good'' jet.

Higgs physics is, however, not the only inspiration behind the
proposed study of these ditau objects.  Viewed in the light of
recently observed anomalies in the $B_s \rightarrow J/\psi\, \phi$
decay~\cite{Aaltonen:2007he} and the like-sign dimuon charge
asymmetry~\cite{Abazov:2010hv}, the search for light ditau resonances
is especially important: New physics solutions to these anomalies
often modify the physics of $B_s-\bar{B}_s$ mixing, and result in a
large decay width of $B_s$ to $\tau^{+}\tau^{-}$~\cite{btautau} or
predict new ditau resonances that mix with $B_s$ mesons~\cite{bsmix}.
Whether it is used for determining the $B_s$ decay width to ditaus, or for
discovering a new light ditau resonance at the LHC, an analysis with
ditau jets becomes a powerful tool in the search for new physics.

A collinear ditau configuration at considerably larger transverse
momentum is also present in scenarios such as $Z'\rig h Z, h\rig
\tau^+ \tau^-$, when the $Z'$ has a mass in the multi-TeV region,
while $m_h\sim 120\gev$. This scenario has been studied in detail in
Ref.~\cite{Katz:2010iq}, whose authors relied on extremely hard jets and
found their analysis to be useful only with a much bigger event sample
($ \sim 100~\text{fb}^{-1}$).  
We, on the contrary, seek a strategy
that can isolate even moderately hard ditau jets and is relevant for
the initial LHC run ($ \sim 10~\text{fb}^{-1}$).

In this work, we consider the NMSSM light pseudoscalar $a$ to be the
candidate for $\mathcal{A}$ and a source for the ditau
jets\footnote{In the NMSSM for large $\tan\beta$ and $m_a \lesssim
  10~\text{GeV}$ this is the dominant decay mode and an important
  search channel \cite{Carena:2007jk}.}.  We focus on the topology in
Eq.~\eqref{eq:ditaudec}.  Since our primary interest lies in tagging
ditau jets, we assume, without loss of generality, that the branching
fraction of $h\to \mathcal{A}\mathcal{A}$ is $100\%$. For a more
realistic and model dependent case study, our results need to be
rescaled with the corresponding branching fractions.

The rest of this work is organized as follows: We introduce and
discuss observables which differentiate ditau jets from QCD light and
heavy flavor jets at low and moderate $p_T$ in
Section~\ref{sec:jetcomp}.  Subsequently, in
Section~\ref{sec:analysis}, we show that there are realistic chances
of discovering the Higgs boson from a sample of $pp \rig Z h+X$ events
when the differentiating power of these observables is used along with
more traditional cuts and requirements. In Section~\ref{sec:tagger},
we first propose a ditau tagger after combining these ditau
discriminating observables to a likelihood, and then comment on the
prospects and efficiency of its tagging performance in a broader
context. Finally, in Section~\ref{sec:conc} we provide our concluding
remarks.

%%%%%%%%%%%%%%%%%%%%%%%%%%%
%%%%%%%%%%%%%%%%%%%%%%%%%%%
%%%% SECTION
%%%%%%%%%%%%%%%%%%%%%%%%%%%
%%%%%%%%%%%%%%%%%%%%%%%%%%%
\section{Ditau jets vs. QCD jets}
\label{sec:jetcomp}
\begin{figure}[!t]
  \includegraphics[height=0.34\textwidth]{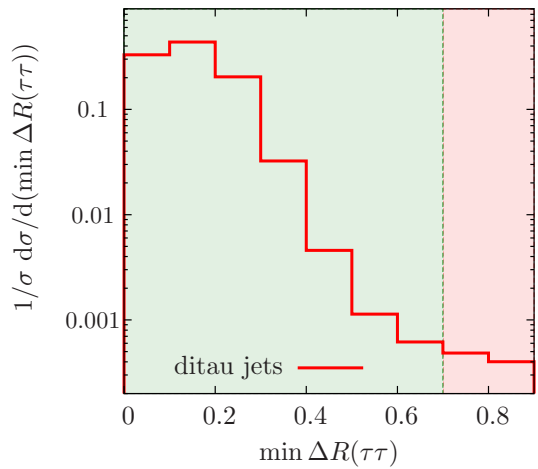}
  \caption{\label{fig:rtautau} Minimum ditau separation of the
    considered signal event sample. Vertical line is the cone size of
    $R=0.7$ jet and roughly gives the jet cone size depending on the
    jet's transverse momentum.}
\end{figure}

The ditau jet -- a jet consisting of a couple of collinear tau
particles, is visibly different from an ordinary QCD jet in terms of
the pattern of energy deposited in the calorimeters. We devote this
section to quantifying these differences in terms of various kinematic
observables. We use {\sc{Pythia}}~\cite{Sjostrand:2006za} generated
$pp\rig hZ$ events, where the Higgs decays to a couple of ditau
resonances, namely, $\mathcal{A}$.  These are the source of our ditau
jets.  To be more specific, we choose
\begin{equation}
  \label{eq:mass-setup}
  m_h=120\gev \qquad \text{and} \qquad 
  m_{\mathcal{A}}=10\gev\,.
\end{equation}
The lightness of the ditau resonance ensures the collinearity of the
resultant tau particles. In Figure~\ref{fig:rtautau} we show the
minimum ditau separation in the azimuthal angle-pseudorapidity plane
for the above event sample.  Since jets constructed with the
jet-algorithm parameter $R$ typically contain particles separated by
$\Delta R \leq R$, a jet with $R = 0.7$ and made out of these events
almost always includes the entire decay products of the ditau
resonance.

We differentiate these ditau jets from jets constructed out of
{\sc{Pythia}} generated dijet event samples. We separately generate
and compare dijets of light flavor ($u, d, s$ and $g$), and heavy
flavors ($c$ and $b$).

%%%%%%%%%%%%%%%%%%%%%%%%%%
%%%%%%%%%%%%%%%%%%            FIGURE
%%%%%%%%%%%%%%%%%%%%%%%%%%
%%%%%%%%%%%%%%%%%%%%%%%%%%
%%%%%%%%%%%%%%%%%%            FIGURE
\begin{figure*}[!t]
  \includegraphics[height=0.34 \textwidth]{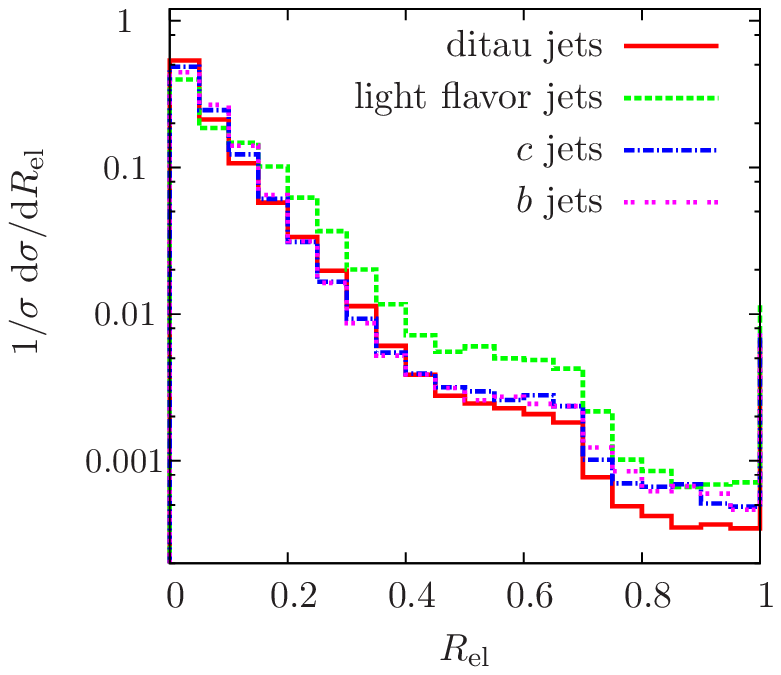}
  \hspace{2cm}
  \includegraphics[height=0.34 \textwidth]{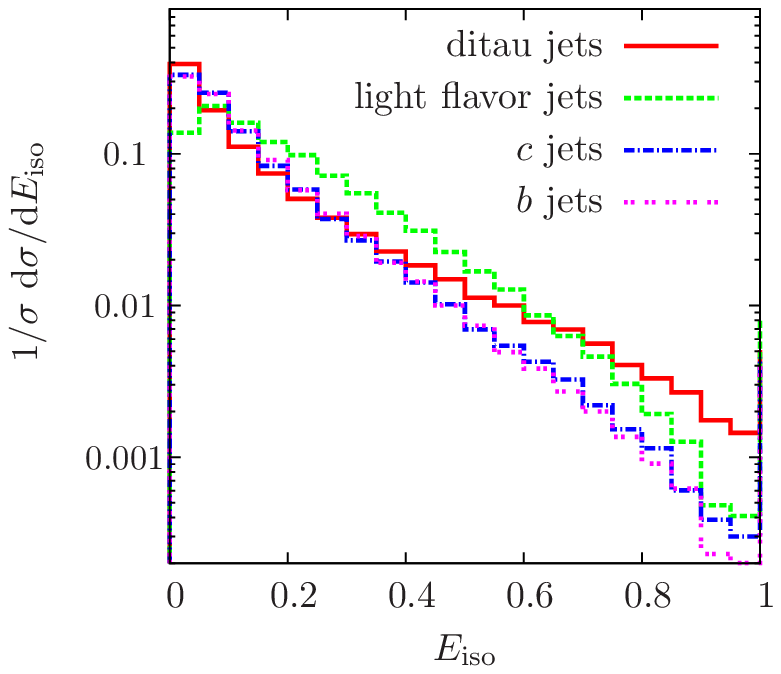}
  \caption{\label{fig:tauobs} Distributions of the electromagnetic
    radius $R_{\text{el}}$ and the energy isolation $E_{\text{iso}}$,
    Eqs.~\gl{eq:rel},~\gl{eq:eiso}. To determine $E_{\text{iso}}$ we 
    choose $r_1=0.2$ and $r_2=0.4$ for illustration purposes and we
    emphasize that our results are qualitatively unaltered for different $r_1,r_2$
    choices. }
%  \vspace{-0.3cm}
\end{figure*}
%%%%%%%%%%%%%%%%%%%%%%%%%%

%%%%%%%%%%%%%%%%%%%%%%%%%%%
%%%% SUBSECTION
%%%%%%%%%%%%%%%%%%%%%%%%%%%
\subsection{Technical setup}
Throughout this work, we include finite detector resolution effects.
We analyze events with a hybrid electromagnetic/hadronic calorimeter
(ecal/hcal) as implemented in {\sc{Pgs}} \cite{pgs}. More concretely,
we construct massless four-vectors separately from hits in the ecal
and hcal grids with thresholds $E=0.5\gev$ and granularities
\begin{align}
  \text{ecal : }\quad  &  \Delta\eta \times \Delta \phi = 0.025 \times 0.025\,,\\
  \text{hcal : }\quad& \Delta\eta \times \Delta \phi = 0.1 \times 0.1\,,
\end{align}
where $\eta$ and $\phi$ denote pseudorapidity and azimuthal angle
respectively.  This allows us to access ecal and hcal information
separately in the analysis. Note that this is an important handle in
constructing observables such as the ecal/hcal energy ratios and the
energy distribution within the jet, typically
employed when discriminating tau jets from QCD jets
\cite{exitau,elusive}.

Muons and electrons are reconstructed from their ecal four-vectors and
their MC-generated energies. For the purpose of our analysis we are
predominantly interested in the light leptons' four-momenta
granularized on the ecal grid. In the actual experiment it is the
combination of calorimeter entries and tracking information which
allows precise reconstruction of the light leptons' four momenta. Here
we implicitly assign the total energy of the lepton, as determined
from the above combined measurement, to the ecal hit. We define an
electron or a muon to be isolated if the hadronic energy deposit
within a cone of size $R=0.3$ is smaller than $10\%$ of the lepton
candidate's transverse momentum.

Jets are constructed out of the rest of the massless four vectors. In
particular, we use the anti-$k_T$ algorithm with $R=0.7$ as implemented in
{\sc{FastJet}}~\cite{Cacciari:2005hq}. 

%%%%%%%%%%%%%%%%%%%%%%%%%%
%%%%%%%%%%%%%%%%%%            FIGURE
\begin{figure*}[!t]
 \includegraphics[height=0.34\textwidth]{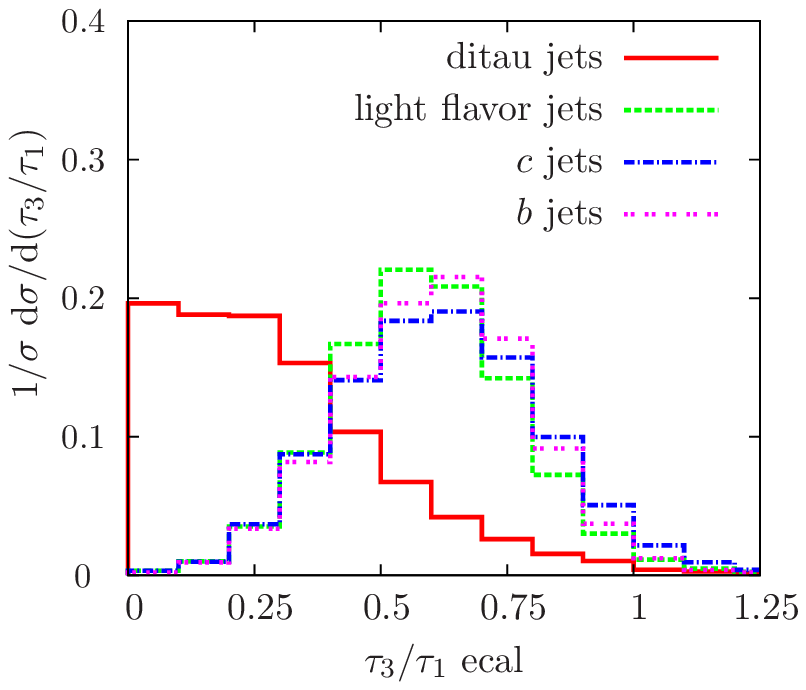}
 \hspace{2cm}
 \includegraphics[height=0.34\textwidth]{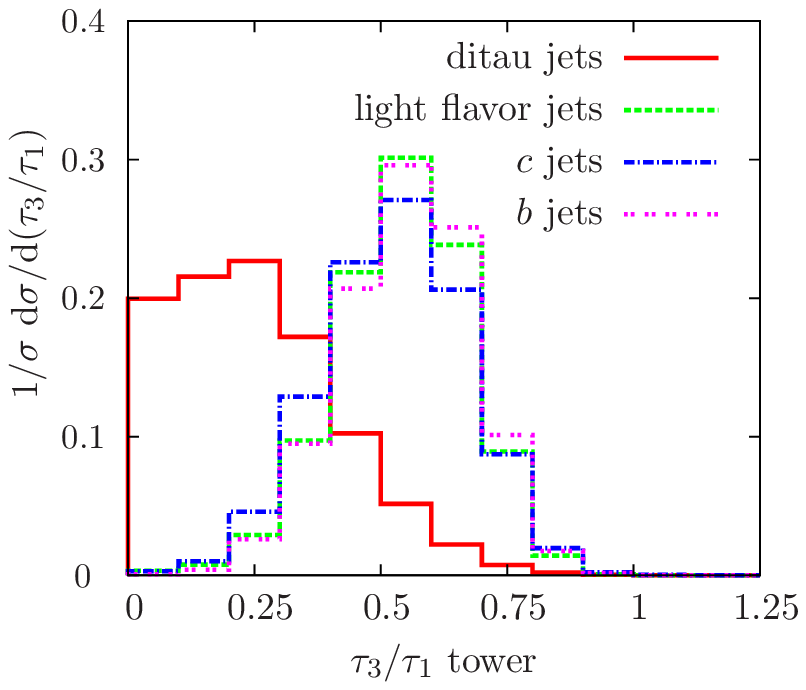}
 \caption{\label{fig:subjetti} Normalized differential distributions of the
   $N$-subjettiness ratio $\tau_3/\tau_1$. In the left panel we plot
   this ratio for ecal hits only, and in the right panel we plot 
  $\tau_3/\tau_1$ for the full calorimeter tower entries.}
    \label{fig:nsub}
    \vspace{-0.3cm}
\end{figure*}
%%%%%%%%%%%%%%%%%%%%%%%%%%
%%%%%%%%%%%%%%%%%%%%%%%%%%
%%%%%%%%%%%%%%%%%%            FIGURE
\begin{figure*}[!t]
  \includegraphics[height=0.33\textwidth]{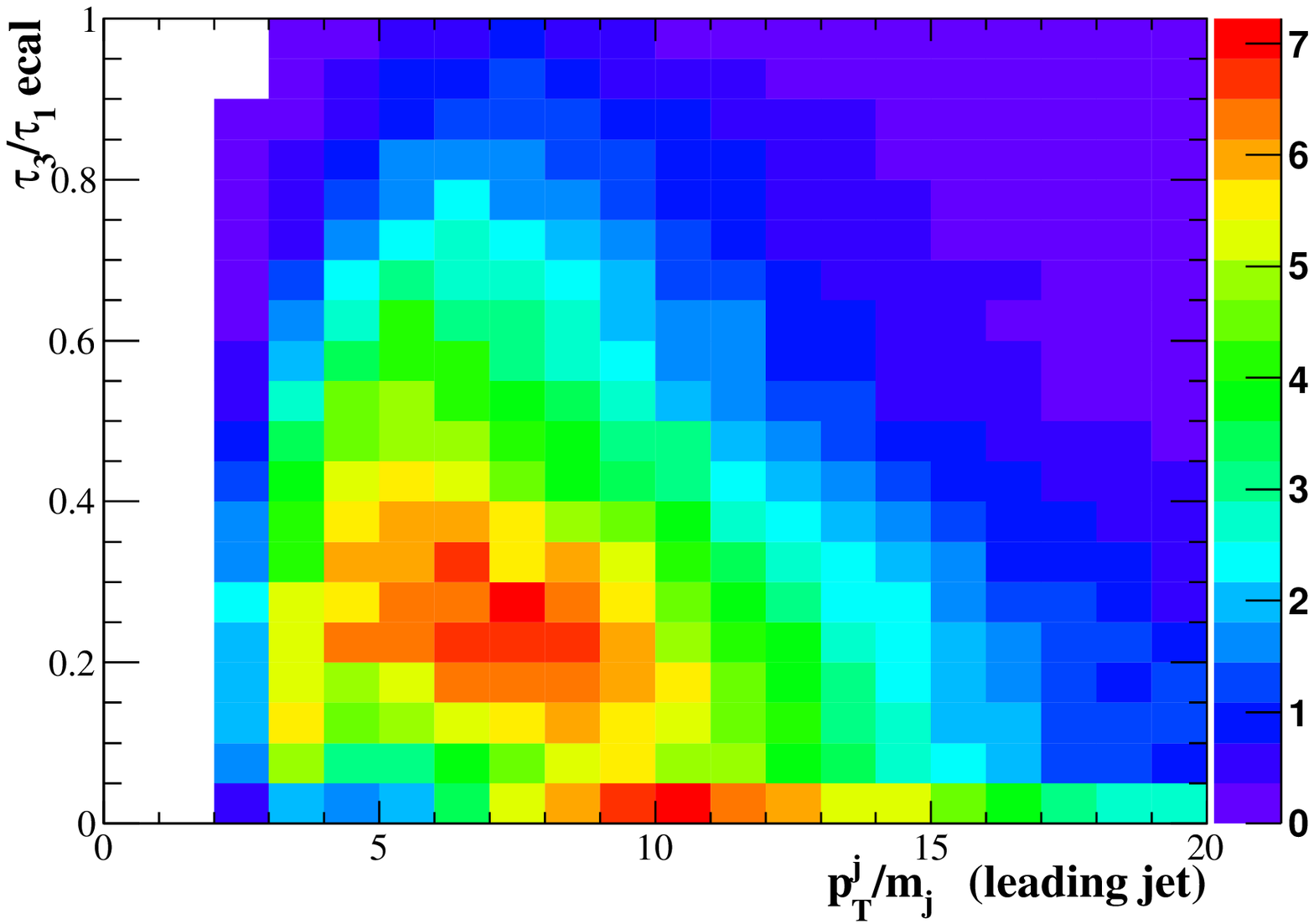}
  \includegraphics[height=0.33\textwidth]{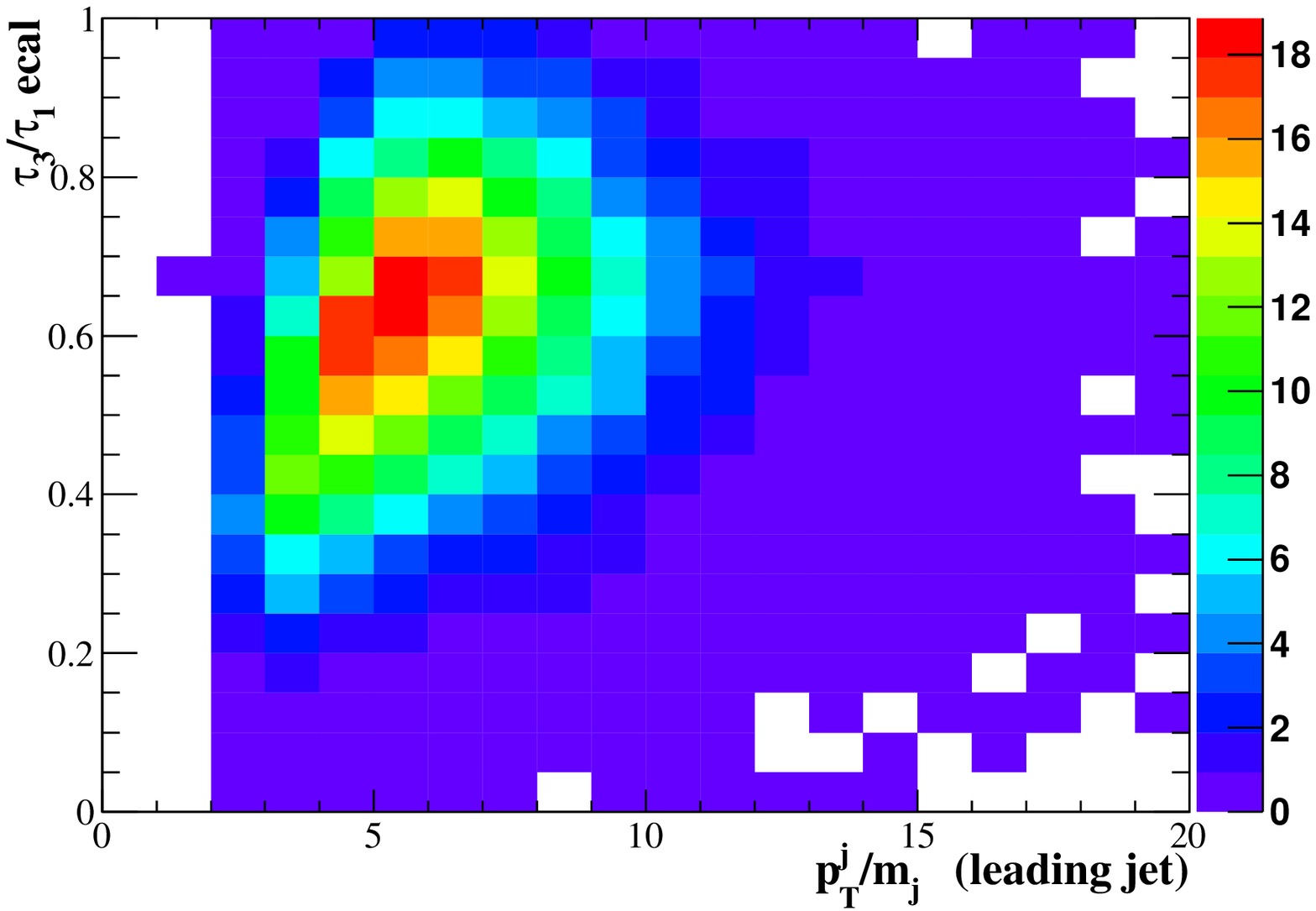}
  \caption{\label{fig:2dnsub} Two-dimensional distribution of $10^3$
    signal events (left panel) and $10^3$ $b$ jet events in the
    $\tau_3/\tau_1$--$p_T^j/m_j$ plane. The ditau events are less
    correlated than QCD jets.}
\end{figure*}
%%%%%%%%%%%%%%%%%%%%%%%%%%

%%%%%%%%%%%%%%%%%%%%%%%%%%%
%%%% SUBSECTION
%%%%%%%%%%%%%%%%%%%%%%%%%%%
\subsection{Discriminating ditau jets}
\label{sec:ditaujets}
Usually $\tau$ decays are classified in so-called `$n$-pronged'
decays, where `$n$' specifies the number of isolated charged tracks
associated with the $\tau$-jet.  Even for a ditau jet, the associated
number of charged tracks still remains a powerful differentiator. In
this work, however, we emphasize the prongness of the energy deposited
in the calorimeters to isolate a ditau jet.

Let us first note that the various decay modes of the tau particle can
be summarized as follows~\cite{Nakamura:2010zzi}:
\begin{subequations}
\label{eq:taudecays}
  \begin{equation}
  \begin{split}
    \tau^\pm \rig e^\pm,\mu^\pm + \slashed{p}_T &\quad  35\%\,, \\
    \tau^\pm \rig \text{hadrons} + \slashed{p}_T &\quad 65\%\,,
    \end{split}
  \end{equation}
which is tantamount to ditau jet branching ratios
\begin{equation}
  \begin{split}
    \text{ditaus decay leptonically} &\quad  12.25\%\,, \\
    \text{ditaus decay semi-leptonically} &\quad  45.5\%\,, \\
    \text{ditaus decay hadronically} &\quad 42.25\%\,.
   \end{split}
\end{equation}
\end{subequations}

Naively one expects that the leptonic or the semi-leptonic decay
channels of the ditau resonance can easily be tagged due to one or
more associated leptons~\cite{Katz:2010iq}. However, that is not the
case for a moderately hard ditau resonance, which only gives rise to
soft leptons. Similar decay patterns are also observed in the case of $B$
and $D$ mesons and a tagging algorithm based on identifying these soft
leptons would give rise to large fake rates. In our analysis, we treat
all decay modes listed in Eq.~\gl{eq:taudecays} on an equal footing.

Before introducing new variables, let us first show that the
traditional calorimeter based algorithms for identifying tau-jets are
not that potent as far as tagging a ditau jet is concerned.  To do
this, we consider the electromagnetic radius
\begin{equation}
  \label{eq:rel}
  R_{\text{em}}^j=\sum_\alpha p_{T,\alpha}\,\Delta R(\alpha,j) 
  \bigg/ \sum_\alpha p_{T,\alpha} \,,
\end{equation}
and the jet energy isolation
\begin{equation}
  \label{eq:eiso}
  E^j_{\text{iso}} = \sum_{r_1\leq \Delta R(\alpha,j) \leq r_2} 
  p_{T,\alpha}  \bigg/ \sum_\alpha p_{T,\alpha} \,,
\end{equation}
associated with a jet $j$. Here the index $\alpha$ runs over only the
ecal cells of the jet, and $\Delta R(\alpha, j)$ is the angular
distance of the $\alpha-$th ecal cell from the jet. Note that both
these quantities enter the tau-jet discriminating likelihood of
Ref.~\cite{exitau} and play crucial roles in tagging a tau jet.  As
shown in Figure~\ref{fig:tauobs}, ditau jets do not show
sufficiently different profiles from ordinary QCD jets in either of
these distributions. Consequently, it is evident that a naive
application of single tau strategies to ditau jets results in a bad
tagging performance.

Both $R_{\text{el}}$ and $E_{\text{iso}}$ are designed to find a clean
jet, i.e. a jet where most of the energy is deposited in only a
few calorimeter cells that are also in close proximity to each other.
A tau jet is such a jet since the hadronic decay products of the tau
arise from a color-singlet state and the decay itself is an
electroweak process.  Although, we expect less radiation in a ditau
jet compared to QCD~\cite{Gallicchio:2010sw}, it is still not a clean
jet. Unlike QCD jets, a ditau jet deposits energy in the calorimeter
in a prong-like fashion -- a feature which we exploit in our analysis.
More specifically, we consider
$N$-subjettiness~\cite{kim,Thaler:2010tr}, a derivative of the
recently proposed observable $N$-jettiness~\cite{Stewart:2010tn}. We
use the definition of~\cite{Thaler:2010tr}
\begin{equation} 
\label{eq:subjetti}
  \tau_N = \frac{\sum_k p_{T,k}\min\left(\Delta
      R(1,k),\dots ,\Delta R(N,k)\right)}{  \sum_j
    p_{T,j}\,R }\,,
\end{equation}
where the indices $k,j$ runs over the fat jet-constituents and the
index $N$ denotes the number of required subjets. Note that in this
definition, $\Delta R(i,k)$ denotes the distance from the jet
constituent $k$ to the subjet $i$. $N$-subjettiness\footnote{We do not
  perform a minimization procedure to retrieve a global event shape
  observable, but instead use the exclusive $k_T$ algorithm as
  implemented in {\sc{FastJet}} on the jet's constituents to cluster
  exactly $N$ subjets.} is particularly successful in discriminating
those jets that have a substructure of isolated collimated energy deposits
from jets that have a more fanned out substructure. Actually, the
ratios between different $\tau_N$ are found to be superior to plain
$\tau_N$ distributions in discriminating jets with multiprong
structures~\cite{Thaler:2010tr}.  In the case of ditau jets, we find that
$\tau_3/\tau_1$, whether of the full calorimeter tower or of the ecal
entries, works best (see Figure~\ref{fig:nsub}). It must be emphasized
that in order to get small values of $\tau_N$, $N$ does not need to
match the number of charged decay products of the taus; rather it must
match those of the pronounced energy deposits in the jet.

\begin{figure}[!t]
  \includegraphics[height=0.34\textwidth]{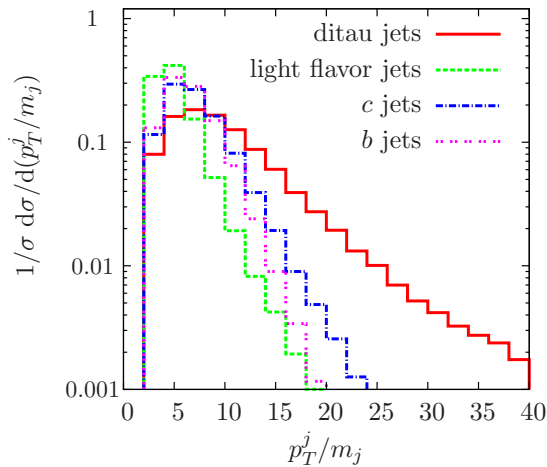}
  \caption{\label{fig:ptjmj} Normalized differential distribution of
    the ratio $p_T^j/m_j$ of the leading jet of the event. For details
    see the text.}
    \label{fig:ptm}	
\end{figure}

We find the ratio between a jet's transverse momentum and mass,
$p_T^j/m_j$, to be another powerful discriminator between a ditau jet
and an ordinary QCD jet. This quantity is sensitive to the size of the
"active area" of the jet (the area where radiation is measured), and
to the alignment between hard radiation and jet axis. For the
pencil-like structure of the ditau jets we expect larger values of
$p_T^j/m_j$ than for QCD jets, which we confirm in
Figure~\ref{fig:ptm}.

More interestingly, a distinct pattern of correlation is observed in
the $N$-subjettiness vs. $p_T^j/m_j$ plane for the ditau jets. For the
QCD jets, as shown in Figure~\ref{fig:2dnsub}, increasing values of
$\tau_3/\tau_1$ are correlated with increasing $p_T^j/m_j$, whereas for
ditau jets these are anti-correlated.

Before concluding this section let us also briefly mention that the
number of charged tracks associated with a jet can play a crucial role
in tagging ditau jets. In Figure~\ref{fig:chargedtracks} we have
plotted the distribution of the number of charged tracks with $p_T\geq
2\gev$.  As expected, a much larger fraction of ditau jets contain $2$
or fewer tracks in them.

%%%%%%%%%%%%%%%%%%%%%%%%%%
%%%%%%%%%%%%%%%%%%            FIGURE
%%%%%%%%%%%%%%%%%%%%%%%%%%

%%%%%%%%%%%%%%%%%%%%%%%%%%%
%%%%%%%%%%%%%%%%%%%%%%%%%%%
%%%% SECTION
%%%%%%%%%%%%%%%%%%%%%%%%%%%
%%%%%%%%%%%%%%%%%%%%%%%%%%%
\section{Ditau jets at work: a case study}
\label{sec:analysis}
As a concrete phenomenological example, we apply the ditau sensitive
observables from Sec.~\ref{sec:jetcomp} to extract Higgses out of
$pp\rig h(\mathcal{ A A}) \, Z (\ell^+ \ell^-)+X $ events via
identifying ditau jets (candidates for $\mathcal{A}$ particles).  For
a realistic discovery potential, however, we also need to be concerned
about the rates. Consequently, we employ the discriminating powers of
only $\tau_3/\tau_1$ and $p_T^j/m_j$ to isolate ditau jets after
imposing realistic pre-selection requirements.

We again use the mass parameters quoted in Eq.~\gl{eq:mass-setup} and
assume $\text{BR}(\mathcal{A}\rightarrow \tau^+ \tau^-) = 100\%$.  The
signal events are characterized by $(i)$ two isolated hard leptons of
identical flavor that reconstruct a $Z$ boson, $(ii)$ a couple of
ditau jets, and $(iii)$ a sizable amount of missing energy.
Therefore, the biggest backgrounds to the signal events are due to
$ZZj$, $WZj$, $WWj$ and $t\bar t$ events. Note that $Z(\ell^+
\ell^-)$+jets can also be an important background, since jet
mismeasurement can give rise to a finite $\slashed{p}_T$.  On the
other hand, $Z/\gamma$+jets actually are SM-candle processes for
getting a handle on these detector and jet energy-scale induced
effects in a data-driven approach (see, e.g., Ref.~\cite{cmsmet}).  It
is impossible to realistically asses these systematic effects without
carrying out the full detector simulation, especially because we are
explicitly looking into the phase space region where the missing
energy tends to be aligned with the leading jet (see
below). Therefore, we have to rely on the experiments to estimate this
contribution to the background. Given that the jet energy scale can be
determined at the level of a few percent \cite{jetscale}, we believe
that this background can be reliably reduced and we do not include
these events in our background analysis.

We produce the matched diboson+jet samples with
{\sc{Sherpa}}~\cite{Gleisberg:2008ta}. The QCD corrections to these
processes have been provided recently in Ref.~\cite{Campbell:2007ev}
(see also Ref.~\cite{Campanario:2010xn} for further details on
precision diboson+jet phenomenology).  We find the $K$ factors
associated with $W^-Zj$ ($W^+Zj$) production to be $0.825 (0.884)$
using {\sc{Vbfnlo}}~\cite{Arnold:2008rz}, i.e. the QCD corrections
reduce the leading order results in the considered phase space before
subjets cuts are imposed.  For the $ZZj$ and $WWj$ production, there
are currently no publicly available codes. However, it is known that
the QCD corrections are qualitatively and qualitatively
similar~\cite{Campbell:2007ev,Campanario:2010xn}, and we adopt the $K$
factor for $WZj$ events for the remaining $VVj$ backgrounds.

%%%%%%%%%%%%%%%%%%%%%%%%%%
%%%%%%%%%%%%%%%%%%            FIGURE
\begin{figure}[!t]
  \includegraphics[height=0.34\textwidth]{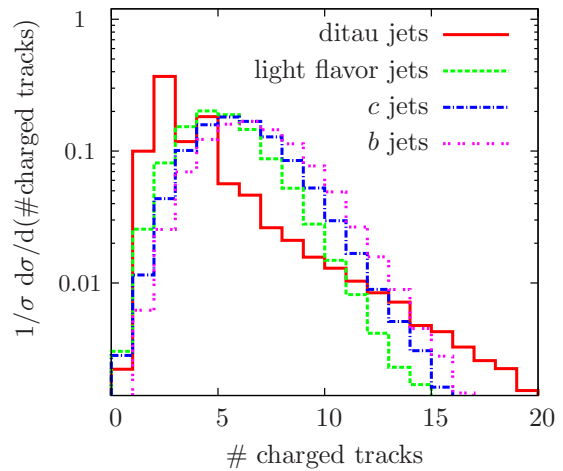}
  \caption{\label{fig:tracks} Normalized differential distribution of
    the number of charged tracks for ditau jets and QCD jets.}
    \label{fig:chargedtracks}	
\end{figure}
%%%%%%%%%%%%%%%%%%%%%%%%%%
%

The $t\bar{t}$ sample is produced using
{\sc{Herwig++}}~\cite{Bahr:2008pv}; we normalize to the NLO QCD cross
section of $815$ pb~\cite{Cacciari:2008zb}. The ditau signal events
are generated using {\sc{Pythia}}.

Even a very large ditau-jet tagging efficiency is still too low to
compete with the large $VVj$ and $t\bar t$ backgrounds on an inclusive
level if we also take into account a small mistagging probability. We,
therefore, apply a number of $S/B$-improving pre-selection criteria
before we can use the ditau sensitive observables of the previous
section. A similar strategy to reduce the backgrounds has been used
in~\cite{Batell:2011tc}. The cuts are summarized in
Table~\ref{tab:cutflow} and we describe them in detail as follows:
%%%%%%%%%%%%%%%%%%%%%%%%%%
%%%%%%%%%%%%%%%%%%            TABLE
\begin{table*}
\begin{tabular}{l | c c c c c}
& ditaus & $ZZj$ & $WZj$ & $WWj$ & $t\bar t$ \\ 
\hline
\hline
& 1.00 & 1.00 & 1.00 & 1.00 & 1.00 \\
\hline
$n_\ell=2,$ &  \multirow{2}{*}{0.416} & \multirow{2}{*}{0.217} & \multirow{2}{*}{0.130}  & \multirow{2}{*}{0.011} & \multirow{2}{*}{0.026} \\ 
$Z$ mass reconstruction with $e^+e^-$ or $\mu^+ \mu^-$ \\
\hline
$ {\text{max}}\,(p^\ell_T,p^{\ell'}_T)\geq 80~{\text{GeV}},~
p_T^Z\geq 150~{\text{GeV}}$  & 0.216 & 0.048 & 0.035  & 0.00019 & $3.9~10^{-4}$ \\
\hline
$n_j\geq 1~{\text{with}}~p_T^j\geq 30~{\text{GeV}},~{\text{no}}~\Delta R(j_{50},Z)\leq 1.5$
& 0.199 &  0.0402 & 0.029 & 0.00019 & $3.0~10^{-4}$ \\
\hline
$\slashed{p}_T\geq 50 ~{\text{GeV}},~|\Delta\phi ({\slashed{\vec{p}}},Z)|\geq 2 $ 
& 0.172 & 0.033 & 0.021 & 0.00015  & $4.6~10^{-5}$ \\
\hline
$\tau_3/\tau_1 |_{\text{ecal}} \leq 0.5$ (leading jet) & 0.125  & 0.011& 0.0084 & $5.4~10^{-5}$ & $2.1~10^{-5}$\\
\hline
$p_T^j/m_j\geq 7$ (leading jet)  & 0.083 & 0.0018 &  0.0020 & $3.0~10^{-6}$ & $ 7.2~10^{-6}$\\
\hline
\hline
%cross section [fb]	& 1.32 & 0.51 & 2.08 & 0.21  & 0.29 \\
cross section [fb]	& 1.32 & 0.45 & 1.83 & 0.18  & 0.29 \\
% k factor for VVj = 0.88
\hline
\hline
\end{tabular} 
\caption{\label{tab:cutflow}
Acceptances for the different steps of the analysis described in
Sec.~\ref{sec:analysis}. The last row gives the cross sections
after all steps have been carried out, including the $K$ factors from
QCD corrections (for details see the text).}
\end{table*}
%%%%%%%%%%%%%%%%%%%%%%%%%%

\begin{enumerate}
\item First, we require exactly two isolated leptons
  (cf.~Sec.~\ref{sec:jetcomp}) of identical flavor, each with $p_T
  \geq 30\gev$. The leptons should reconstruct the $Z$ mass within
  $10\gev$. This ensures that the event is triggerable and reduces the
  large $WW+\text{jets}$ and $t\bar{t}$ backgrounds.
\item We impose staggered $p_T$ criteria on the two leptons and focus
  on the boosted kinematics with the Higgs recoiling against the $Z$
  in the signal sample, requiring $p_{T}^{Z}>150\gev$.
\item We require at least one $R=0.7$ anti-$k_T$ jet with a transverse
  momentum $p_T \geq 30\gev$.  We veto events that contain hard jets
  ($p_T^j\geq 50\gev$) close to the reconstructed $Z$ ($\Delta
  R(\text{jet}, Z)\leq 1.5$). We apply this cut to reduce the large
  $t\bar{t}$ background.  In $t\bar{t}$, to fake the boosted $Z$ boson
  and pass the staggered $p_{T}^\ell$ cuts, one of the tops need to be
  boosted and its decay products need to be collimated. The veto
  removes $t\bar{t}$ events where the $b$ quark is close to the hard
  lepton.
\item We require $\slashed{p}_T\geq 50\gev$. The reconstructed missing
  transverse momentum vector additionally needs to be separated from
  the reconstructed $Z$ by $|\Delta\phi(\slashed{\vec{p}}_T,Z)|\geq
  2$. The $\tau$ decays always produce neutrinos which result in a
  sizable amount of missing transverse energy. Together with the cuts
  on the lepton system, $\slashed{p}_T$ provides a good handle against
  the large $Z+\text{jets}$ background.
\end{enumerate}
In the two final steps of the analysis we apply the results from
Sec.~\ref{sec:jetcomp} in a rectangular fashion:

%%%%%%%%%%%%%%%%%%%%%%%%%%
%%%%%%%%%%%%%%%%%%            FIGURE
\begin{figure}[!t]
  \includegraphics[height=0.34\textwidth]{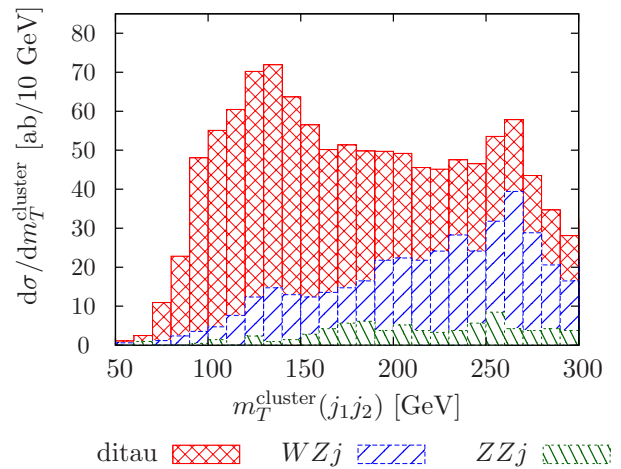}
  \caption{\label{fig:hmassdis} Transverse cluster mass distribution
    after all cuts of TABLE \ref{tab:cutflow} have been applied
    with the additional requirement of having at least two jets.  Not
    shown are the completely suppressed $WWj$ and $t\bar t$
    backgrounds.}
\end{figure}
%%%%%%%%%%%%%%%%%%%%%%%%%%
\begin{enumerate}
  \setcounter{enumi}{4}
\item We require $\tau_3/\tau_1<0.5$ on the ecal level, and
\item the leading jet is characterized by \hbox{$p_T^j/m_j\geq 7$}.
\end{enumerate}

The two final steps are crucial in increasing the ditau signal over
the contributing backgrounds. In total this leaves us with
\begin{equation}
S/B = 0.48
\end{equation}
and a signal cross section after cuts of
\begin{equation}
\sigma(\text{signal after cuts})=1.32 \fb \,. 
\end{equation}

This means that we can achieve $S/\sqrt{B}\gtrsim 5$ for
\hbox{${\cal{L}} \simeq 40\ifb$}. There is obviously a lot of
parameter space left to relax our assumptions\footnote{For completeness, 
we note that there are certain parameter regions where the Higgs
does have a sizable branching ratio to four taus~\cite{Dermisek:2010mg}.}
$\text{BR}(h\to{\cal{A}}{\cal{A}})=\text{BR}({\cal{A}}\to \tau^+\tau^-)=100\%$.
While this result is already good enough to constrain this specific
model class, reconstructing the Higgs mass is more challenging.  This
is not a straightforward task since, due to the multiple tau decays,
the missing transverse momentum's direction $\slashed{\vec{p}}_T$ is
largely unconstrained. We impose an additional requirement of having
at least two jets with $p_T^j\geq 30\gev$
($\sigma({\text{signal}})=0.73~{\text{fb}}$) and calculate the
so-called transverse cluster mass~\cite{mtcluster}
\begin{multline}
  [m^{\text{cluster}}_{T}(j_1j_2)]^2= \left( \sqrt{m^{2}(j_1j_2)+p_{T}^{2}(j_1j_2}) +
  |\slashed{p}_{T}| \right)^2 \\ - \left( {\vec{p}}_{T}(j_1j_2) +\slashed{\vec{p}}_{T}
  \right)^2\,. 
\end{multline}
We, hereby, do not impose the two final selection criteria of
Table~\ref{tab:cutflow} on the next-to-leading jet since we find that
the next-to-leading jet may also be a single-tau jet or can be due to
initial state radiation. Applying the ditau criteria reduces the
signal cross section, but, in principle improves the mass resolution.
We stress that the signal and the background have completely different
$m_T^{\text{cluster}}$ shapes and that the bulk of the signal cross
section clusters around the Higgs mass of $120\gev$. While the
distributions are likely to be altered by detector effects, one can
clearly use the low $m_T^{\text{cluster}}$ region to constrain or even
measure a possible low-$m_T^{\text{cluster}}$ excess if the model is
realized.

Note, reconstructing the mass of the Higgs is generically troublesome
if the Higgs particle decays to partly invisible final states. In
addition to that, hadronic observables which in principle allow to
reconstruct the Higgs excess are heavily affected by initial state
radiation and underlying event. The later also affects the definition
of the transverse cluster mass since it explicitly depends on the jet
mass. Further, initial state radiation can produce one of the two
hardest jets in the event. As a result, the jacobian peak gets washed
out significantly as shown in Figure~\ref{fig:hmassdis} (compared to
the good resolution in purely leptonic final states as considered in
Ref.~\cite{mtcluster}).  Nonetheless, side-band analyses seem very
promising. When restricting $m^{\text{cluster}}_{T}(j_1j_2)<160\gev$
we find $\sigma({\text{signal}})=0.50~{\text{fb}}$ and
$\sigma({\text{background}})=0.12~{\text{fb}}$, which yields
$S/\sqrt{B}\gtrsim 5$ for ${\cal{L}}= 12\ifb$.

%%%%%%%%%%%%%%%%%%%%%%%%%%%%
%%%%%%%%%%%%%%%%%%%%%%%%%%%%
%%%%% SECTION
%%%%%%%%%%%%%%%%%%%%%%%%%%%%
%%%%%%%%%%%%%%%%%%%%%%%%%%%%
\section{Toward low $p_T$ ditau tagging}
\label{sec:tagger}

In this section we combine the (sub)jet observables of
Sec.~\ref{sec:ditaujets} to a likelihood,
\begin{equation}
  \label{eq:likelihood}
  L=f\left(\tau_3/\tau_1 |_{\text{ecal}}\right)\times f ( p^j_T/m_j )  \times f(
  \text{charged~tracks})
\end{equation}
where the $f(.)$ is the probability distribution of the respective
observable in Figures~\ref{fig:subjetti} and \ref{fig:ptjmj}.  In
Eq.~\gl{eq:likelihood} we have also included the number of charged
tracks distribution, which adds additional discriminative power on top
of $\tau_3/\tau_1$ and $p_T^j/m_j$ according to
Figure~\ref{fig:tracks}.

From this likelihood we can construct a single quantity $d$ by a
standard procedure (an exercise similar to that is done for
$b$-tagging~\cite{exibtag}), which discriminates ditau jets from light
flavor, $c$ and $b$ jets,
\begin{multline}
  \label{eq:singledis}
  d=p(\text{light~flavor})\,{L({\text{ditau}}) \over L({\text{ditau}}) +  L({\text{light flavor}})}\\
  + p(\text{$c$})\,{L({\text{ditau}}) \over L({\text{ditau}}) +
    L({\text{$c$}})} + p(\text{$b$})\,{L({\text{ditau}}) \over
    L({\text{ditau}}) + L({\text{$b$}})}\,.
\end{multline}
The function $p(.)$ denotes the a priori probability of having a light
flavor jet, a $c$ jet, or a $b$ jet. Therefore,
$p(\text{light~flavor})+p(c)+p(b)=1$. We choose these probabilities by
counting the color and flavor degrees and completely disregard the
parton distributions in the initial states:
\begin{equation}
  \label{eq:apriori}
  p(\text{$c$})=p(\text{$b$})=3/23\,,\quad p(\text{light~flavor})=17/23\,.
\end{equation}
Since the distributions of the QCD jets are less sensitive to the
flavor content of the jets, this choice has only a small impact on the
actual distribution of $d$.  The result of the choice in
Eq.~\gl{eq:apriori} is shown in Fig.~\ref{fig:combdis}. Considering
jets with $p_T\geq 30\gev$ with $d>0.7$ gives a ditau-tagging
efficiency of $66\%$ ($58\%$) and with an average mistagging
probability of $7\%$ ($6\% $ ) if charged tracks are included (not
included).

%%%%%%%%%%%%%%%%%%%%%%%%%%
%%%%%%%%%%%%%%%%%%            FIGURE
\begin{figure}[!t]
  \includegraphics[height=0.34\textwidth]{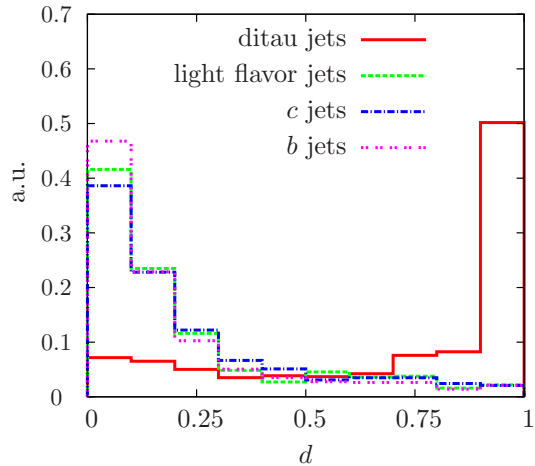}
  \caption{\label{fig:combdis} Combined discriminator,
    Eq.~\gl{eq:singledis}, that results from the likelihood of
    Eq.~\gl{eq:likelihood}.}
    \vspace{0.5cm}
\end{figure}
%%%%%%%%%%%%%%%%%%%%%%%%%%
%%%%%%%%%%%%%%%%%%%%%%%%%%
%%%%%%%%%%%%%%%%%%            FIGURE
\begin{figure}[!t]
  \includegraphics[height=0.34\textwidth]{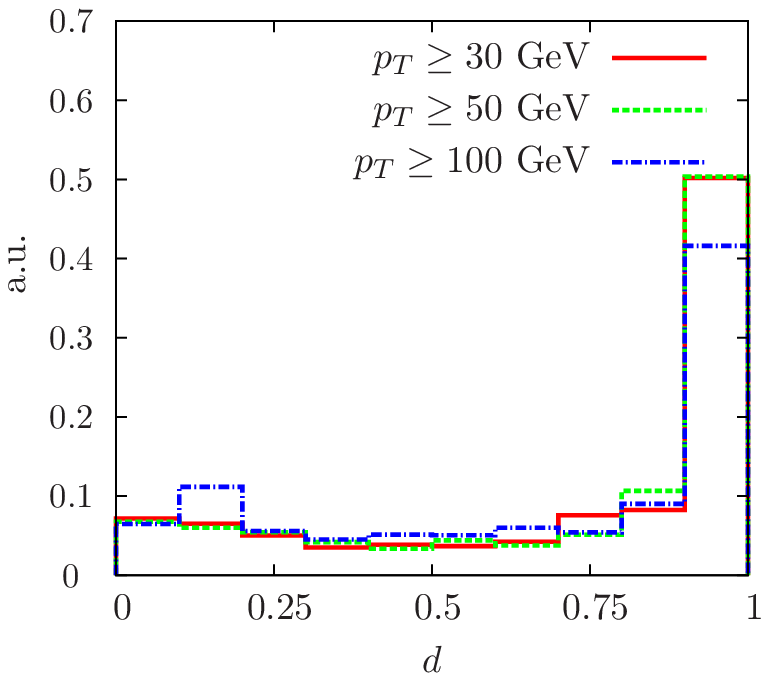}\\[0.6cm]
  \includegraphics[height=0.34\textwidth]{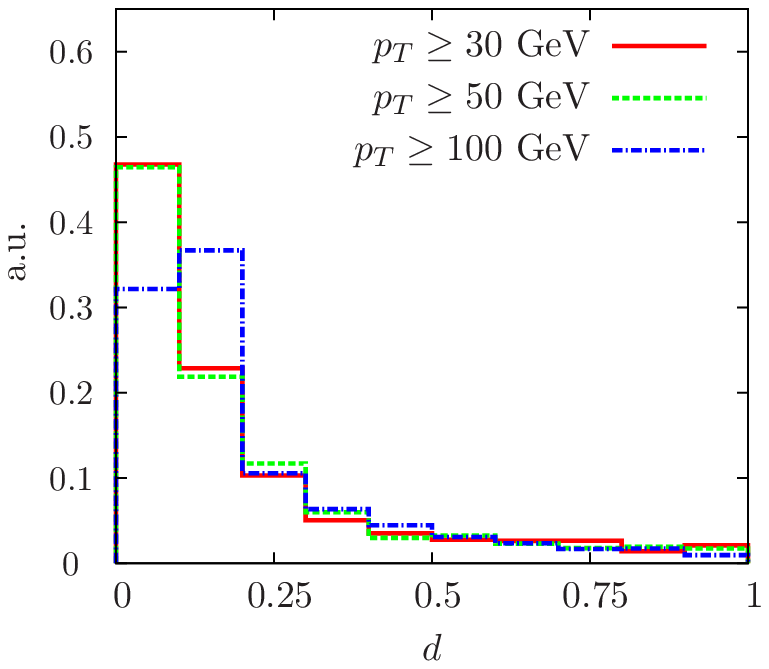}
  \caption{\label{fig:combdiscom} Combined discriminators for $b$ jets
    (lower panel) and ditau jets (upper panel) for different minimum
    transverse momentum requirements on the leading jet.}
\end{figure}
%%%%%%%%%%%%%%%%%%%%%%%%%%

The tagging efficiency is, of course, a function of the considered
jet's transverse momentum as shown in Figure~\ref{fig:combdiscom}. For
larger transverse momenta, $p_T^j/m_j$ looses its discriminative
power, while the discriminative features of the $\tau_3/\tau_1$
observable remain intact.

%%%%%%%%%%%%%%%%%%%%%%%%%%%
%%%%%%%%%%%%%%%%%%%%%%%%%%%
%%%% SECTION
%%%%%%%%%%%%%%%%%%%%%%%%%%%
%%%%%%%%%%%%%%%%%%%%%%%%%%%
\section{Summary and Conclusions}
\label{sec:conc}

Non-standard Higgs sectors with non-standard Higgs decays require
dedicated analysis strategies in order not to miss evidence of new
physics when analyzing early LHC data. In this letter we have argued, that
straightforwardly applying tau recognition algorithms to jets which
actually consist of a boosted tau pair does not lead to a satisfactory
signal-over-background discrimination. Consequently, current analysis
strategies are not suitable to cope with these signatures which
might well arise in scenarios of strong interactions, the NMSSM or in
hidden-valley-type models.

We have shown that the combination of only two observable,
$N$-subjettiness (in particular $\tau_3/\tau_1$) and $p_T^j/m_j$ does
serve to highly lift the degeneracy of ditau jets and QCD jets of all
kind. The combination of both observables encodes orthogonal
information on the ditaus' distinct radiation and decay pattern.

We have applied these results in a phenomenological analysis of
$pp\rig hZ+X,~Z\rig \ell^+\ell^-,~\text{BR}(h \rig
\mathcal{A}\mathcal{A})=\text{BR}(A\to \tau^+\tau^-)=100\%$ for
boosted kinematics and have shown that constraints can be formulated
for small integrated luminosities and that the Higgs mass for the
chosen parameters can, in principle, be reconstructed. More
specifically, we find that the Higgs signal becomes statistically
significant for luminosities ${\cal{L}}\simeq 12\ifb$ leaving enough
space to compensate smaller branching ratios in more realistic
scenarios.

Constructing a toy ditau tagger based on the two observables,
augmented by the number of charged tracks distribution in a likelihood
approach, we find a high tagging efficiency with an acceptably small
mistagging probability, not too sensitive on the considered jet's
$p_T$.

Of course, our results are subject to modifications when confronted
with all contributing experimental uncertainties and mass setups
different from Eq.~\gl{eq:mass-setup}.  However, our findings strongly
motivate a more detailed investigation within a full detector simulation
framework.\bigskip

{\bf{Acknowledgments}} --- 
%\section*{Acknowledgments}
CE thanks Bob McElrath and Tilman Plehn for discussions and Jeannette
Bloch-Ditzinger for all the enjoyable coffee breaks. TSR thanks Anna
Goussiou for conversations. CE and MS thank the Fermilab Theoretical
Physics Department for hospitality during the time when parts of this
work were completed.

The simulations underlying this study have been performed in parts on
bwGRiD (\url{http://www.bw-grid.de}), member of the German D-Grid
initiative, funded by the Ministry for Education and Research
(Bundesministerium f\"ur Bildung und Forschung) and the Ministry for
Science, Research and Arts Baden-W\"urttemberg (Ministerium f\"ur
Wissenschaft, Forschung und Kunst Baden-W\"urttemberg). We thank
Daniel Steck and Jeremy Thorn for the use of the Quantum Control
computer cluster at the University of Oregon.  MS was supported by the
US Department of Energy under contract number DE-FG02-96ER40969.  TSR
was supported in part by the US Department of Energy under contract
number DE-FGO2-96ER40956.

%%%%%%%%%%%%%%%%%%%%%%%%%%%%
%%%%%%%%%%%%%%%%%%%%%%%%%%%%
%%%%%%%%%%%%%%%%%%%%%%%%%%%%
%%%%%%%%%%%%%%%%%%%%%%%%%%%%
%%%%%%%%%%%%%%%%%%%%%%%%%%%%


\begin{thebibliography}{99}

\bibitem{Aad:2008zzm}
  G.~Aad {\it et al.}  [ATLAS Collaboration],
  %``The ATLAS Experiment at the CERN Large Hadron Collider,''
  JINST {\bf 3} (2008) S08003.
  %%CITATION = JINST,3,S08003;%%
  
\bibitem{Ball:2007zza}
  G.~L.~Bayatian {\it et al.}  [CMS Collaboration],
  %``CMS technical design report, volume II: Physics performance,''
  J.\ Phys.\ G {\bf 34} (2007) 995.
  %%CITATION = JPHGB,G34,995;%%

\bibitem{Gao:2010qx}
  C.~P.~Buszello, I.~Fleck, P.~Marquard and J.~J.~van der Bij,
  %``Prospective Analysis Of Spin- And Cp-Sensitive Variables In H $\to$ Z Z
  %$\to$ L(1)+ L(1)- L(2)+ L(2)- At The Lhc,''
  Eur.\ Phys.\ J.\  C {\bf 32} (2004) 209;
  %  [arXiv:hep-ph/0212396].
  %%CITATION = EPHJA,C32,209;%%
  Y.~Gao, A.~V.~Gritsan, Z.~Guo, K.~Melnikov, M.~Schulze and N.~V.~Tran,
  %``Spin determination of single-produced resonances at hadron colliders,''
  Phys.\ Rev.\  D {\bf 81}, 075022 (2010);
  %  [arXiv:1001.3396 [hep-ph]].
  %%CITATION = PHRVA,D81,075022;%%  
  A.~De Rujula, J.~Lykken, M.~Pierini, C.~Rogan and M.~Spiropulu,
  %``Higgs look-alikes at the LHC,''
  Phys.\ Rev.\  D {\bf 82} (2010) 013003;
  %  [arXiv:1001.5300 [hep-ph]].
  %%CITATION = PHRVA,D82,013003;%%
  C.~Englert, C.~Hackstein and M.~Spannowsky,
  %``Measuring spin and CP from semi-hadronic ZZ decays using jet
  %substructure,''
  Phys.\ Rev.\  D {\bf 82} (2010) 114024.
  %  [arXiv:1010.0676 [hep-ph]]
  %%CITATION = PHRVA,D82,114024;%%

\bibitem{Butterworth:2008iy}
  J.~M.~Butterworth, A.~R.~Davison, M.~Rubin and G.~P.~Salam,
  %``Jet substructure as a new Higgs search channel at the LHC,''
  Phys.\ Rev.\ Lett.\  {\bf 100} (2008) 242001.
  %  [arXiv:0802.2470 [hep-ph]].
  %%CITATION = PRLTA,100,242001;%%

\bibitem{Soper:2010xk}
  D.~E.~Soper, M.~Spannowsky,
  %``Combining subjet algorithms to enhance ZH detection at the LHC,''
  JHEP {\bf 1008}, 029 (2010).
  %  [arXiv:1005.0417 [hep-ph]].
  %%CITATION = JHEPA,1008,029;%%

\bibitem{MSSM}
  G.~D.~Kribs, A.~Martin, T.~S.~Roy and M.~Spannowsky,
  %``Discovering the Higgs Boson in New Physics Events using Jet Substructure,''
  Phys.\ Rev.\  D {\bf 81} (2010) 111501;
  % [arXiv:0912.4731 [hep-ph]].
  %%CITATION = PHRVA,D81,111501;%%
  G.~D.~Kribs, A.~Martin, T.~S.~Roy and M.~Spannowsky,
  %``Discovering Higgs Bosons of the MSSM using Jet Substructure,''
  Phys.\ Rev.\  D {\bf 82} (2010) 095012
  %[arXiv:1006.1656 [hep-ph]].
  %%CITATION = PHRVA,D82,095012;%%

\bibitem{Tprime}
  G.~D.~Kribs, A.~Martin and T.~S.~Roy,
  %``Higgs Discovery through Top-Partners using Jet Substructure,''
  arXiv:1012.2866 [hep-ph].
  %%CITATION = ARXIV:1012.2866;%%

\bibitem{exitau}
  M.~Heldmann, D.~Cavalli,~
  ATL-PHYS-PUB-2006-008, ATL-COM-PHYS-2006-010.

\bibitem{Plehn:1999xi}
  D.~L.~Rainwater, D.~Zeppenfeld and K.~Hagiwara,
  %``Searching for H ---> tau tau in weak boson fusion at the CERN LHC,''
  Phys.\ Rev.\  D {\bf 59} (1998) 014037;
  %[arXiv:hep-ph/9808468].
  %%CITATION = PHRVA,D59,014037;%%
  T.~Plehn, D.~L.~Rainwater and D.~Zeppenfeld,
  %``A Method for identifying H ---> tau tau ---> e+- mu-+ p(T) at the CERN
  %LHC,''   Phys.\ Rev.\  D {\bf 61} (2000) 093005;
  %[arXiv:hep-ph/9911385].
  %%CITATION = PHRVA,D61,093005;%%
  B.~Mellado, W.~Quayle and S.~L.~Wu,
  %``Prospects for the observation of a Higgs boson with H ---> tau+ tau- --->
  %l+l- p-slash(t) associated with one jet at the LHC,''
  Phys.\ Lett.\  B {\bf 611} (2005) 60.
  %  [arXiv:hep-ph/0406095].

  
\bibitem{Martin:2011pd}
  A.~Martin and T.~S.~Roy,
  %``The Gold-Plated Channel for Supersymmetric Higgs via Higgsphilic Z',''
  arXiv:1103.3504 [hep-ph].
  %%CITATION = ARXIV:1103.3504;%%

\bibitem{Dermisek:2005ar}
  R.~Dermisek and J.~F.~Gunion,
  %``Escaping the large fine tuning and little hierarchy problems in the next to
  %minimal supersymmetric model and h ---> aa decays,''
  Phys.\ Rev.\ Lett.\  {\bf 95} (2005) 041801;
%  [arXiv:hep-ph/0502105].
  %%CITATION = PRLTA,95,041801;%%
  R.~Dermisek, J.~F.~Gunion and B.~McElrath,
  %``Probing NMSSM Scenarios with Minimal Fine-Tuning by Searching for Decays of
  %the Upsilon to a Light CP-Odd Higgs Boson,''
  Phys.\ Rev.\  D {\bf 76} (2007) 051105;
%  [arXiv:hep-ph/0612031].
  R.~Dermisek and J.~F.~Gunion,
  %``New constraints on a light CP-odd Higgs boson and related NMSSM Ideal Higgs
  %Scenarios,''
  Phys.\ Rev.\  D {\bf 81} (2010) 075003.
%  [arXiv:1002.1971 [hep-ph]].
  %%CITATION = PHRVA,D81,075003;%%

  %%CITATION = PHRVA,D76,051105;%%
    
\bibitem{Maniatis:2009re}
  M.~Maniatis,
  %``The Next-to-Minimal Supersymmetric extension of the Standard Model
  %reviewed,''
  Int.\ J.\ Mod.\ Phys.\  A {\bf 25} (2010) 3505;
  %  [arXiv:0906.0777 [hep-ph]].
  %%CITATION = IMPAE,A25,3505;%%
   U.~Ellwanger, C.~Hugonie and A.~M.~Teixeira,
  %``The Next-to-Minimal Supersymmetric Standard Model,''
  Phys.\ Rept.\  {\bf 496} (2010) 1;
  %  [arXiv:0910.1785 [hep-ph]].
  %%CITATION = PRPLC,496,1;%%
  S.~Chang, P.~J.~Fox, N.~Weiner,
  %``Naturalness and Higgs decays in the MSSM with a singlet,''
  JHEP {\bf 0608}, 068 (2006).
  %  [hep-ph/0511250].
  %%CITATION = JHEPA,0608,068;%%

\bibitem{Giudice:2007fh}
  G.~F.~Giudice, C.~Grojean, A.~Pomarol and R.~Rattazzi,
  %``The Strongly-Interacting Light Higgs,''
  JHEP {\bf 0706} (2007) 045;
  %  [arXiv:hep-ph/0703164].
  %%CITATION = JHEPA,0706,045;%%
  B.~Bellazzini, C.~Csaki, A.~Falkowski and A.~Weiler,
  %``Buried Higgs,''
  Phys.\ Rev.\  D {\bf 80} (2009) 075008.
  %  [arXiv:0906.3026 [hep-ph]].
  %%CITATION = PHRVA,D80,075008;%%
  
\bibitem{compsearch}
  R.~Contino, C.~Grojean, M.~Moretti, F.~Piccinini and R.~Rattazzi,
  %``Strong Double Higgs Production at the LHC,''
  JHEP {\bf 1005} (2010) 089;
  %  [arXiv:1002.1011 [hep-ph]].
  %%CITATION = JHEPA,1005,089;%%
  J.~R.~Espinosa, C.~Grojean and M.~Muhlleitner,
  %``Composite Higgs Search at the LHC,''
  JHEP {\bf 1005} (2010) 065; 
  %  [arXiv:1003.3251 [hep-ph]].
  %%CITATION = JHEPA,1005,065;%%
  A.~Falkowski, D.~Krohn, L.~T.~Wang, J.~Shelton and A.~Thalapillil,
  %``Unburied Higgs,''
  arXiv:1006.1650 [hep-ph];
  %%CITATION = ARXIV:1006.1650;%%
  R.~Grober and M.~Muhlleitner,
  %``Composite Higgs Boson Pair Production at the LHC,''
  arXiv:1012.1562 [hep-ph].
  %%CITATION = ARXIV:1012.1562;%%

\bibitem{Schabinger:2005ei}
  R.~Schabinger and J.~D.~Wells,
  %``A Minimal spontaneously broken hidden sector and its impact on Higgs boson
  %physics at the large hadron collider,''
  Phys.\ Rev.\  D {\bf 72} (2005) 093007;
  %[arXiv:hep-ph/0509209].
  %%CITATION = PHRVA,D72,093007;%%
  B.~Patt and F.~Wilczek,
  %``Higgs-field portal into hidden sectors,''
  arXiv:hep-ph/0605188;
  %%CITATION = HEP-PH/0605188;%%
  M.~J.~Strassler and K.~M.~Zurek,
  %``Echoes of a hidden valley at hadron colliders,''
  Phys.\ Lett.\  B {\bf 651}, 374 (2007).
  %[arXiv:hep-ph/0604261].
  %%CITATION = PHLTA,B651,374;%%
  
\bibitem{hiddenpheno}
   S.~Bock, R.~Lafaye, T.~Plehn, M.~Rauch, D.~Zerwas and P.~M.~Zerwas,
  %``Measuring Hidden Higgs and Strongly-Interacting Higgs Scenarios,''
  Phys.\ Lett.\  B {\bf 694}, 44 (2010);
  %  [arXiv:1007.2645 [hep-ph]].
  %%CITATION = PHLTA,B694,44;%%
  C.~Englert, T.~Plehn, D.~Zerwas and P.~M.~Zerwas,
  %``Exploring the Higgs portal,''
  arXiv:1106.3097 [hep-ph].
  %%CITATION = ARXIV:1106.3097;%%

\bibitem{Aaltonen:2007he}
  T.~Aaltonen {\it et al.}  [CDF Collaboration],
  %``First Flavor-Tagged Determination of Bounds on Mixing-Induced CP Violation
  %in $B^0_{s} \to J/\psi \phi$ Decays,''
  Phys.\ Rev.\ Lett.\  {\bf 100} (2008) 161802;
  %  [arXiv:0712.2397 [hep-ex]].
  %%CITATION = PRLTA,100,161802;%%
    V.~M.~Abazov {\it et al.}  [D0 Collaboration],
  %``Measurement of $B^0_{s}$ mixing parameters from the flavor-tagged decay
  %$B^0_{s} \to J/\psi \phi$,''
  Phys.\ Rev.\ Lett.\  {\bf 101} (2008) 241801;
  %  [arXiv:0802.2255 [hep-ex]].
  %%CITATION = PRLTA,101,241801;%%
  A.~Lenz {\it et al.},
  %``Anatomy of New Physics in $B - \bar{B}$ mixing,''
  Phys.\ Rev.\  D {\bf 83} (2011) 036004.
  %  [arXiv:1008.1593 [hep-ph]].
  %%CITATION = PHRVA,D83,036004;%%

\bibitem{Abazov:2010hv}
  V.~M.~Abazov {\it et al.}  [D0 Collaboration],
  %``Evidence for an anomalous like-sign dimuon charge asymmetry,''
  Phys.\ Rev.\  D {\bf 82} (2010) 032001.
  %  [arXiv:1005.2757 [hep-ex]].
  %%CITATION = PHRVA,D82,032001;%%


\bibitem{btautau}
  A.~Dighe, A.~Kundu and S.~Nandi,
  %``Enhanced $B_s - \bar{B}_s$ lifetime difference and anomalous like-sign
  %dimuon charge asymmetry from new physics in $B_s -> \tau^+ \tau^-$,''
  Phys.\ Rev.\  D {\bf 82} (2010) 031502.
  %  [arXiv:1005.4051 [hep-ph]].
  %%CITATION = PHRVA,D82,031502;%%


\bibitem{bsmix}
  Y.~Bai and A.~E.~Nelson,
  %``CP Violating Contribution to Delta Gamma in the B_s System from Mixing with
  %a Hidden Pseudoscalar,''
  Phys.\ Rev.\  D {\bf 82} (2010) 114027.
  %  [arXiv:1007.0596 [hep-ph]].
  %%CITATION = PHRVA,D82,114027;%%

\bibitem{Katz:2010iq}
  A.~Katz, M.~Son and B.~Tweedie,
  %``Ditau-Jet Tagging and Boosted Higgses from a Multi-TeV Resonance,''
  arXiv:1011.4523 [hep-ph].
  %%CITATION = ARXIV:1011.4523;%%

\bibitem{Carena:2007jk}
  M.~Carena, T.~Han, G.~Y.~Huang and C.~E.~M.~Wagner,
  %``Higgs Signal for h $\to$ aa at Hadron Colliders,''
  JHEP {\bf 0804}, 092 (2008).
  %  [arXiv:0712.2466 [hep-ph]].
  %%CITATION = JHEPA,0804,092;%%

\bibitem{Sjostrand:2006za}
  T.~Sjostrand, S.~Mrenna and P.~Z.~Skands,
  %``PYTHIA 6.4 Physics and Manual,''
  JHEP {\bf 0605}, 026 (2006);
  \url{http://home.thep.lu.se/~torbjorn/Pythia.html}.
  %  [arXiv:hep-ph/0603175].
  %%CITATION = JHEPA,0605,026;%%

\bibitem{pgs}
   J.~Conway {\it et al.},
  \url{http://www.physics.ucdavis.edu/~conway/research/software/pgs/pgs.html}.

\bibitem{elusive}
  C.~R.~Chen, M.~M.~Nojiri and W.~Sreethawong,
  %``Search for the Elusive Higgs Boson Using Jet Structure at LHC,''
  JHEP {\bf 1011} (2010) 012.
  %  [arXiv:1006.1151 [hep-ph]].
  %%CITATION = JHEPA,1011,012;%% 

\bibitem{Cacciari:2005hq}
  M.~Cacciari and G.~P.~Salam,
  %``Dispelling the $N^{3}$ myth for the $k_t$ jet-finder,''
  Phys.\ Lett.\  B {\bf 641}, 57 (2006);
  %  [arXiv:hep-ph/0512210];
  M. Cacciari, G. P. Salam and G. Soyez, {\url{http://fastjet.fr}}.
  %%CITATION = PHLTA,B641,57;%%

\bibitem{Nakamura:2010zzi}
  K.~Nakamura {\it et al.}  [Particle Data Group],
  %``Review of particle physics,''
  J.\ Phys.\ G {\bf 37} (2010) 075021.
  %%CITATION = JPHGB,G37,075021;%%

\bibitem{Gallicchio:2010sw}
  G.~Marchesini, B.~R.~Webber,
  %``Simulation of QCD Jets Including Soft Gluon Interference,''
  Nucl.\ Phys.\  {\bf B238}, 1 (1984);
  %%CITATION = NUPHA,B238,1;%%
  J.~Gallicchio and M.~D.~Schwartz,
  %``Seeing in Color: Jet Superstructure,''
  Phys.\ Rev.\ Lett.\  {\bf 105} (2010) 022001;
  %  [arXiv:1001.5027 [hep-ph]].
  %%CITATION = PRLTA,105,022001;%%
  A.~Hook, M.~Jankowiak and J.~G.~Wacker,
  %``Jet Dipolarity: Top Tagging with Color Flow,''
  arXiv:1102.1012 [hep-ph];
  %%CITATION = ARXIV:1102.1012;%%
  D.~E.~Soper, M.~Spannowsky,
  %``Finding physics signals with shower deconstruction,''
  arXiv:1102.3480 [hep-ph].
  %%CITATION = ARXIV:1102.3480;%%
  
\bibitem{kim}
  J.~H.~Kim,
  %``Rest Frame Subjet Algorithm With SISCone Jet For Fully Hadronic Decaying
  %Higgs Search,''
  Phys.\ Rev.\  D {\bf 83} (2011) 011502.
  %  [arXiv:1011.1493 [hep-ph]].
  %%CITATION = PHRVA,D83,011502;%%

\bibitem{Thaler:2010tr}
  J.~Thaler and K.~Van Tilburg,
  %``Identifying Boosted Objects with N-subjettiness,''
  JHEP {\bf 1103} (2011) 015.
  %  [arXiv:1011.2268 [hep-ph]].
  %%CITATION = JHEPA,1103,015;%%

 \bibitem{Stewart:2010tn}
  I.~W.~Stewart, F.~J.~Tackmann, W.~J.~Waalewijn,
  %``N-Jettiness: An Inclusive Event Shape to Veto Jets,''
  Phys.\ Rev.\ Lett.\  {\bf 105}, 092002 (2010).
  %  [arXiv:1004.2489 [hep-ph]].  
  
\bibitem{cmsmet}
The CMS Collaboration, CMS PAS SUS-10-001;
The CMS collaboration, CMS-PAS-JME-09-005;
The Atlas collaboration, ATL-PHYS-PUB-2009-015.
 
\bibitem{jetscale}
 see, e.g., The Atlas collaboration, ATLAS-CONF-2010-056.

 
\bibitem{Gleisberg:2008ta}
  T.~Gleisberg, S.~Hoeche, F.~Krauss, M.~Schonherr, S.~Schumann, F.~Siegert and J.~Winter,
  %``Event generation with SHERPA 1.1,''
  JHEP {\bf 0902}, 007 (2009);
  %  [arXiv:0811.4622 [hep-ph]],
  %%CITATION = JHEPA,0902,007;%%
  S.~Schumann and F.~Krauss,
  %``A Parton shower algorithm based on Catani-Seymour dipole factorisation,''
  JHEP {\bf 0803} (2008) 038; 
  %  [arXiv:0709.1027 [hep-ph]],
  %%CITATION = JHEPA,0803,038;%%
  T.~Gleisberg and S.~Hoeche,
  %``Comix, a new matrix element generator,''
  JHEP {\bf 0812} (2008) 039;
  %  [arXiv:0808.3674 [hep-ph]],
  %%CITATION = JHEPA,0812,039;%%
  M.~Schonherr and F.~Krauss,
  %``Soft Photon Radiation in Particle Decays in SHERPA,''
  JHEP {\bf 0812}, 018 (2008),
  \url{http://www.sherpa-mc.de/}.
  %  [arXiv:0810.5071 [hep-ph]].
  %%CITATION = JHEPA,0812,018;%%

\bibitem{Campbell:2007ev}  
  S.~Dittmaier, S.~Kallweit and P.~Uwer,
  %``NLO QCD corrections to WW+jet production at hadron colliders,''
  Phys.\ Rev.\ Lett.\  {\bf 100}, 062003 (2008);
  %  [arXiv:0710.1577 [hep-ph]].
  %%CITATION = PRLTA,100,062003;%%
  J.~M.~Campbell, R.~Keith Ellis and G.~Zanderighi,
  %  ``Next-to-leading order predictions for $WW+1$ jet distributions at the
  %  LHC,''
  JHEP {\bf 0712} (2007) 056;
  %  [{ arXiv:0710.1832}  [hep-ph]].
  %%CITATION = JHEPA,0712,056;%%
  T.~Binoth, T.~Gleisberg, S.~Karg, N.~Kauer and G.~Sanguinetti,
  %``NLO QCD corrections to ZZ+jet production at hadron colliders,''
  Phys.\ Lett.\  B {\bf 683} (2010) 154; 
  %  [arXiv:0911.3181 [hep-ph]].
  %%CITATION = PHLTA,B683,154;%%
  F.~Campanario, C.~Englert, S.~Kallweit, M.~Spannowsky and D.~Zeppenfeld,
  %``NLO QCD corrections to WZ+jet production with leptonic decays,''
  JHEP {\bf 1007} (2010) 076.
  %  [arXiv:1006.0390 [hep-ph]].
  %%CITATION = JHEPA,1007,076;%%

\bibitem{Campanario:2010xn}
  F.~Campanario, C.~Englert, M.~Spannowsky and D.~Zeppenfeld,
  %``NLO-QCD corrections to W gamma j production,''
  Europhys.\ Lett.\  {\bf 88} (2009) 11001;
  %  [arXiv:0908.1638 [hep-ph]].
  %%CITATION = EULEE,88,11001;%%
  F.~Campanario, C.~Englert and M.~Spannowsky,
  %``QCD corrections to non-standard WZ+jet production with leptonic decays at
  %the LHC,''
  Phys.\ Rev.\  D {\bf 82} (2010) 054015;
  %  [arXiv:1006.3090 [hep-ph]].
  %%CITATION = PHRVA,D82,054015;%%
  F.~Campanario, C.~Englert and M.~Spannowsky,
  %``Precise predictions for (non-standard) $W \gamma$ + jet production,''
  Phys.\ Rev.\  D {\bf 83} (2011) 074009.
  %  [arXiv:1010.1291 [hep-ph]].
  %%CITATION = PHRVA,D83,074009;%%

\bibitem{Arnold:2008rz}
  K.~Arnold {\it et al.},
  %``VBFNLO: A Parton level Monte Carlo for processes with electroweak bosons,''
  Comput.\ Phys.\ Commun.\  {\bf 180} (2009) 1661,
  \url{http://www-itp.particle.uni-karlsruhe.de/~vbfnloweb}.
  %  [arXiv:0811.4559 [hep-ph]].
  %%CITATION = CPHCB,180,1661;%%

\bibitem{Bahr:2008pv}
  M.~Bahr {\it et al.},
  %``Herwig++ Physics and Manual,''
  Eur.\ Phys.\ J.\  C {\bf 58} (2008) 639;
  \url{http://projects.hepforge.org/herwig/}.
  %  [arXiv:0803.0883 [hep-ph]].
  %%CITATION = EPHJA,C58,639;%%

\bibitem{Cacciari:2008zb}
  M.~Cacciari, S.~Frixione, M.~L.~Mangano, P.~Nason and G.~Ridolfi,
  %``Updated predictions for the total production cross sections of top and of
  %heavier quark pairs at the Tevatron and at the LHC,''
  JHEP {\bf 0809}, 127 (2008).
  %  [arXiv:0804.2800 [hep-ph]].
  %%CITATION = JHEPA,0809,127;%%

\bibitem{Batell:2011tc}
  B.~Batell, J.~Pradler and M.~Spannowsky,
  %``Dark Matter from Minimal Flavor Violation,''
  arXiv:1105.1781 [hep-ph].
  %%CITATION = ARXIV:1105.1781;%%
  
\bibitem{mtcluster}
 V.~D.~Barger, T.~Han and J.~Ohnemus,
  %``HEAVY LEPTONS AT HADRON SUPERCOLLIDERS,''
  Phys.\ Rev.\  D {\bf 37} (1988) 1174.
  %%CITATION = PHRVA,D37,1174;%%

\bibitem{Dermisek:2010mg}
  R.~Dermisek, J.~F.~Gunion,
  %``New constraints on a light CP-odd Higgs boson and related NMSSM Ideal Higgs Scenarios,''
  Phys.\ Rev.\  {\bf D81 } (2010)  075003;
%  [arXiv:1002.1971 [hep-ph]].  
  F.~Domingo, U.~Ellwanger,
  %``Reduced branching ratio for H -> AA -> 4 tau from A - eta_b mixing,''
  JHEP {\bf 1106 } (2011)  067.
%  [arXiv:1105.1722 [hep-ph]].

\bibitem{exibtag}
C. Weiser,~CMS NOTE 2006/014.

\end{thebibliography}
\end{document}